\documentstyle[aps,prl,psfig]{revtex}

\newcommand{\ket}[1]{{|#1\rangle}}
\newcommand{\bra}[1]{{\langle#1|}}
\newcommand{\trace}{\mbox{tr}}
\renewcommand{\tensor}{\otimes}

\renewcommand{\Re}{\mbox{Re}}
\renewcommand{\Im}{\mbox{Im}}

\newcommand{\cS}{{\cal S}}
\newcommand{\cQ}{{\cal Q}}
\newcommand{\cH}{{\cal H}}

\begin{document}
\twocolumn

\title{On the Power of One Bit of Quantum Information}

\author{E. Knill$^1$, R. Laflamme$^2$}
\address{$^1$ MS B265, $^2$ MS B288, Los Alamos National Laboratory,
Los Alamos, NM 87455}

\date{January 1998}

\maketitle

\begin{abstract}
In standard quantum computation, the initial state
is pure and the answer is determined by making a measurement
of some of the bits in the computational basis.
What can be accomplished if the initial state is
a highly mixed state and the answer is determined by
measuring the expectation of $\sigma_z$ on the first bit with bounded
sensitivity? This is the situation in high temperature
ensemble quantum computation. We show that in this model
it is possible to perform interesting physics simulations
which have no known efficient classical algorithms, even though
the model is less powerful then standard
quantum computing in the presence of oracles.
\end{abstract}


Recent discoveries show that quantum computers can solve problems of
practical interest much faster than known algorithms for classical
computers~\cite{shor:qc1995a,grover:qc1997a}.  This has lead to
widespread recognition of the potential benefits of quantum
computation. Where does the apparent power of quantum computers come
from? This power is frequently attributed to ``quantum parallelism'' ,
interference phenomena derived from the superposition principle, and
the ability to prepare and control pure states according to the
Schr\"odinger equation. Real quantum computers are rarely in pure
states and interact with their environments, which leads to non-unitary
evolution. Furthermore, recent proposals and experiments using NMR at
high temperature to study quantum computation involve manipulations of
extremely mixed states.  Recent research in error-correction and
fault-tolerant computation has shown that non-unitary evolution due to
weak interactions with the environment results in no loss of
computational power, if sufficiently pure states can be
prepared~\cite{knill:qc1998a,preskill:qc1998a,aharonov:qc1996a,kitaev:qc1996a}.
Here we consider the situation where there are no errors or
interactions with the environment, but the initial state is highly
mixed.  We investigate the power of {\em one} bit of quantum
information available for computing, by which we mean that the input
state is equivalent to having one bit in a pure state and arbitrarily
many additional bits in a completely random state. The model of
computation which consists of a classical computer with access to a
state of this form is called {\em deterministic quantum computation
with one quantum bit\/} (DQC1).  We demonstrate that in the presence
of oracles, such a computer is less powerful than one with access to
pure state bits. However, it can solve problems related to physics
simulations for which no efficient classical algorithms are
known. DQC1 is the first non-trivial entry in the class of models of
computations which are between classical computation and standard
quantum computation.  Investigations of such models are expected to
lead to a better understanding of the reasons for the power of quantum
computation.


There are many kinds of problems that one might like to solve using a
computational device. The three main problems not involving
communication are function evaluation, non-deterministic function
evaluation and distribution sampling. Let $\cS$ be the set of all
binary strings and $\cS_n$ the set of binary strings of length
$n$. The $k$'th letter in a binary string $b$ is denoted by $b_k$.  In
function evaluation, we are given a function $f:\cS\rightarrow\cS$ and
a binary string $b$. The goal is to determine $f(b)$.  In
non-deterministic function evaluation, we are given a relation
$R\subseteq\cS\times\cS$ and a binary string $b$. The goal is to
determine a string $c$ such that $R(b,c)$.  In distribution sampling,
we are given a probability distribution $P$ on $\cS$ and a length $n$.
The goal is return a string $b$ of length $n$ sampled according to
$P(b|n)$.

In this paper we focus on deterministic function evaluation.  A
one-bit function is a function whose values are single bits.  Any
deterministic function evaluation problem reduces to that of evaluating
a number of one-bit functions, one for each bit of the output string
(a bound on the length of the output needs to be known).

Any computation requires a certain amount of resources to obtain the
desired answer. The most important resources are time and space. How
much of either is required can depend on the computational power of
the device, which determines the applicable model of computation.  For
a given problem, one usually tries to determine the resources required
as a function of the {\em problem size\/}, which for one-bit functions
is simply the number of bits in the input string. (For general
functions, one can take it to be the sum of the sizes of the input and
the output string.) The computational power of two models of
computation is considered to be the same if for any algorithm in one
model, there is a corresponding algorithm in the other one using at
most a constant multiple of the resources of the first algorithm. An
algorithm is considered to be {\em efficient\/} if the resource
requirements are polynomial in the problem size. For the purpose of
discussing efficient computation, the power of two models is
considered to be the same if for any algorithm in one model, there is
an equivalent algorithm in the other model which uses at most a polynomial
multiple of the resources. We adopt this definition for the remainder
of the paper. In most cases below, the polynomial multiple is linear
or constant. For a comprehensive treatment of classical computational
complexity theory, see~\cite{papdimitriou:qc1994a}.

The models of computation considered here subsume classical
computation. The available computational devices are assumed to
include a classical probabilistic computer conforming (for example) to
the model of an abstract random access machine with access to random
bits, and a quantum system consisting of as many (quantum) bits as
needed.  Without loss of generality, we can assume that the procedure
used to evaluate the function $f$ on input $b$ consists of repetitions
of two parts.  The first uses the classical computer to generate a
sequence of unitary operations. The second applies these operations to
the quantum system in a specified initial state.  A measurement of the
quantum system then yields the answer.  Answers from one repetition
may be used to make decisions in the next ones.  The models differ in
what are the permitted initial states and measurements, and sometimes
in the allowed unitary operations (a.k.a.~quantum gates). In these
models, function evaluation is performed probabilistically and we are
satisfied with an answer which is correct with high probability,
provided that we can make the probability of success high enough.

The state space of the quantum system of $n$ bits is the complex
hilbert space $\cQ^n$ generated by the computational basis. This basis
is labeled by the strings in $\cS_n$; the basis element corresponding
to string $b$ is denoted by $\ket{b}$. $\cQ^n$ is the $n$-fold tensor
product of $\cQ^1$. To describe unitary operators and states, it is
convenient to use the operator basis consisting of tensor products of
the Pauli operators~\cite{somaroo:qc1998a}
\begin{eqnarray}
I&\doteq&\sigma_{00}\doteq
\left[\begin{array}{cc}1&0\\0&1\end{array}\right]\\ 
\sigma_{x}
&\doteq&\sigma_{01}
   \doteq\left[\begin{array}{cc}0&1\\1&0\end{array}\right]\\
\sigma_{y} &\doteq&\sigma_{10}
   \doteq\left[\begin{array}{cc}0&-i\\i&0\end{array}\right]\\
\sigma_{z} &\doteq&\sigma_{11}
   \doteq \left[\begin{array}{cc}1&0\\0&-1\end{array}\right].
\end{eqnarray}
The binary form of the indices consists of two-bit strings. A Pauli
operator acting on the $k$'th bit is denoted by $\sigma^{(k)}_b$. A
general tensor product of Pauli operators is denoted by $\sigma_b$,
where $b_{2k-1}b_{2k}$ is the index of the operator acting on the
$k$'th bit.

A {\em pure state\/} of the quantum system consisting of $n$ bits is a
unit vector in $\cQ^n$. In general, the quantum system can be
correlated with other systems that we do not have access to. As a
result, the general state of the system can be described as a
distribution of pure states. However, this description is
overdetermined and a better one is based on the density operator
formalism.  A state is a positive definite trace one hermitian
operator on $\cQ^n$. Pure states are of the form
$\ket{\psi}\bra{\psi}$ (think of $\ket{\psi}$ as a column vector, and
$\bra{\psi}$ as its hermitian transpose). All other states are
obtained as convex combinations of pure states.

When using highly mixed states, that is those with all eigenvalues
small, it is convenient to describe the state by means of deviations
from the identity expressed as sums of Pauli operators.  In general, a
state $\rho$ can be written in the form
\begin{equation}
\rho = {1\over 2^n}(I+\sum_{b\not={\bf 0}}a_b\sigma_b),
\end{equation}
where the $a_b$ are real and $\mbox{\bf 0}$ is the bit string of all
$0$'s. The deviation of $\rho$ is the non-identity component of the
sum given by $\sum_{b\not={\bf 0}}a_b\sigma_b$.  For example, the
deviation of the state where the first bit is $0$ and the other bits
are completely random is given by $\sigma_z^{(1)}$.

The effect of an operation on the quantum system can be described by
a unitary operator $U$ which maps an input state $\ket{\psi}$
to the output state $U\ket{\psi}$. In terms of operators,
$U$ takes the state $\rho$ to $U\rho U^\dagger$.


In both of the models of quantum computation to be discussed in this
letter, the elementary operations (quantum gates) that can be applied
to the quantum system are a complete set of one and two bit unitary
operators~\cite{divincenzo:qc1995a,barenco:qc1995a}. Since global
phase factors in operations have no effect on the computation, it
suffices to have a complete set of determinant one operators.  A
simple such set consists of $e^{-i\sigma_a^{k}\pi/8}$ for $a\not={\bf
0}$ (the $\pi/8$ rotation) and the controlled $\pi$ rotation
$e^{-i\sigma_z^{k_1}\tensor\sigma_z^{k_2}\pi/4}$~\cite{knill:qc1998a}.
This is a variant of the more familiar conditional sign-flip
operation.  For $k_1=1,k_2=2$, it takes
$\ket{0b}\rightarrow^{-i\sigma_z^{2}\pi/4}\ket{0b}$ and
$\ket{1b}\rightarrow^{i\sigma_z^{2}\pi/4}\ket{1b}$. The difference in
the effect on the second bit depending on the state of the first is a
$\pi$ rotation.  More generally, one can use any $U(b,t)\doteq
e^{-i\sigma_b t}$.  The $U(b,t)$ can be
implemented with fidelity $1-\epsilon$ using $\mbox{\rm
polylog}(1/\epsilon)n$ of the previously mentioned set of operations
(this is a consequence of the results in~\cite{solovay:qc1998a}).  A
network implementation of a unitary operator $U$ is a decomposition of
$U$ as a product of elementary gates, $U=\prod_i G_i$.

\noindent{\bf Deterministic quantum computation with pure states
(DQCp)}: The initial state of the quantum system is $\ket{\mbox{\bf
0}}$ (where bold face means that all bits are in the state $\ket{0}$).  In the
standard model of quantum computation, the final (one-bit) answer is
obtained by a measurement of the first bit. In DQCp, this measurement
is replaced by a process which yields the noisy expectation of
$\sigma_z^{(1)}$ for the final state.  For function evaluation (but
not necessarily the other types of problems), there is no loss of
power by making this restriction on the final measurement.  To be
specific, if the state of the quantum system is $\rho$, the
measurement process returns a number which is sampled from a
distribution with mean
$\langle\sigma_z^{(1)}\rangle=\trace(\sigma_z^{(1)}\rho)$ and bounded
variance $s$, where $s$ is independent of the number of bits used. We
call such a measurement an {\em estimate\/} of the quantity
measured. Repeating the computation and measurement are assumed to
yield independent instances of this distribution.  Thus, the mean can
be estimated to within $\epsilon$ with probability of error at most
$p$ using $O(\log(1/p)/\epsilon^2)$ repetitions of the
computation~\cite{huber:qc1981a}.  DQCp is realized by an idealized
bulk NMR quantum computer, where all molecules are initially perfectly
polarized, there is no decoherence and there is no operational error.

If $U$ is a unitary operator with a network implementation, then DQCp
yields an estimate of $\trace(\sigma_z^{(1)}U\ket{\bf 0}\bra{\bf
0}U^\dagger)$.  By preprocessing the input state with operators of the
form $G=e^{-i\sigma_a\pi/2}$, any quantity of the form
$\trace(\sigma_z^{(1)}U\ket{\bf b}\bra{\bf b}U^\dagger)$ can be
estimated. By further postprocessing the state with one or two
operators of the form $G=e^{-i\sigma_a\pi/4}$, quantities like
$\trace(\sigma_cU\ket{\bf b}\bra{\bf b}U^\dagger)$ can be obtained.
Since the resources required for pre- and postprocessing are linear in
the number of bits, we can identify (the power of) DQCp with the
ability to estimate these quantities.

\noindent{\bf Deterministic quantum computation with  one qubit (DQC1)}:
The deviation of the initial state of the quantum system
is $\sigma_z^{(1)}$. The final answer is obtained as in DQCp by a bounded
variance process yielding $\langle\sigma_z^{(1)}\rangle$.
The initial state corresponds to having one bit
in a pure state and the rest completely random.

DQC1 is realized by an idealized high temperature NMR quantum computer
where there is no decoherence and no operational error. This is the
regime where the initial deviation state can be approximated by
$e^{-\beta H}/Z \sim {1\over 2^n}(I-\beta\sum_ie_i\sigma_z^{(i)})$
(noninteracting bits).  The one-bit initial state can be obtained by
eliminating polarization in bits other than the first. Exploiting the
additional polarization can reduce the time resource required by at
most a factor of $1/n$, so that no additional power can be gained.
Constraints on the amount of polarization extractable from the high
temperature state have been investigated by
Sorensen~\cite{sorensen:qc1989a}.  Note that DQC1 is not robust
against many realistic error-models.  For example, the results of
Aharonov and Ben-Or~\cite{aharonov:qc1996c} show that error-correction
cannot be used to successfully compensate against depolarizing
independent errors due to the rapid loss of information if fresh
ground-state bits cannot be introduced during the computation.

Let $U$ be a unitary operator with a network implementation.  Using
DQC1 and pre- and postprocessing similar to that introduced with DQCp,
we can estimate $\trace({1\over 2^n}\sigma_a U\sigma_b U^\dagger)$ for
any $a$ and $b$. We can therefore identify DQC1 with the ability to
perform these estimates.

DQCp and DQC1 can be used to estimate any coefficient
of certain operator expansions of $U$. For DQC1, write
$U=\sum_b\alpha_b\sigma_b$, with $\alpha_b={1\over 2^n}\trace(\sigma_b U)$.
This is the {\em Pauli operator\/} expansion of $U$.
To determine $\alpha_b$, use the network for $U$ to construct a network
for the operator $V$ which maps $\ket{0}\ket{b}\rightarrow\ket{0}U\ket{b}$
and $\ket{1}\ket{b}\rightarrow\ket{1}\ket{b}$. This is a
``conditional'' $U$ operator and can be implemented
with a linear amount of additional resources~\cite{barenco:qc1995a}. 
Then
\begin{eqnarray}
{1\over 2^{n+1}}\trace(\sigma_{x}^{(1)} V\sigma_{x}^{(1)}\sigma_b V^\dagger)
  &=& {1\over 2}(\trace(U\sigma_b)+\trace(\sigma_b U^\dagger))\\
  &=& \Re(\alpha_b),\nonumber\\
{1\over 2^{n+1}}\trace(\sigma_{y}^{(1)} V\sigma_{x}^{(1)}\sigma_b V^\dagger)
  &=&  {1\over 2}(i\trace(U\sigma_b)-i\trace(\sigma_b U^\dagger))\\
  &=& -\Im(\alpha_b).\nonumber
\end{eqnarray}
Thus the coefficients $\alpha_b$ can be estimated with two computations.
Since $U\sigma_b U^\dagger $ is a unitary operator easily implemented
given networks for $U$, the ability to estimate the coefficients
of the Pauli operator expansion is equal in power to DQC1.

If the trick of the previous paragraph is used with DQCp, we can obtain
any $\trace(U\ket{b}\bra{b})$. Since
$\trace(U\ket{a}\bra{b})$ can be written
as a sum of expressions of the form
$\trace(U{1\over 2}(\ket{a}+e^{i\phi}\ket{b})(\bra{a}+e^{-i\phi}\bra{b})$,
and the states in the second operator can be generated from $\ket{0}$
with $O(n)$ gates, any of the transition amplitudes of $U$
can be determined to get coefficients of the {\em matrix\/} $U$ in the
computational basis.
To see that the ability to estimate $\bra{a}U\ket{b}$ is as powerful
as DQCp, observe that 
$\trace(\sigma_z^{(1)}U\ket{0}\bra{0}U^\dagger) =
\trace(\bra{0}U^\dagger\sigma_z^{(1)} U\ket{0})$.


Let the evolution of a quantum mechanical system be described by a
(possibly time varying) hamiltonian $H$ on a hilbert space $\cH$.  To
efficiently simulate this evolution using a quantum computer with $n$
bits, it is sufficient to have a unitary embedding of $\cH$ in $\cQ^n$
and an extension $H'$ of the embedded hamiltonian for which there are
efficient quantum networks approximating $e^{-iH'(t)\delta}$ (taking
$\hbar = 1$) for small $\delta$ to within $O(\delta^2)$.  (To define
efficiency properly requires problem instances for each $n$, in which
case a resource bound polynomial in $n$ and $1/\delta$ can be
imposed.) A class of such hamiltonians consists of those which are a
polynomially bounded sum of two bit hamiltonians~\cite{lloyd:qc1996a}.
Whether it is useful to implement this simulation depends on what
information one wants to obtain. If one simply wants to predict the
outcome of a measurement in an experiment involving this system, it is
possible to do that by simulation, provided that the initial state can
be computed, and the final measurement can be represented as a
measurement in the computational basis.  DQCp and DQC1 are further
restricted to measurements of expectations of operators which can be
approximated as a sum of a reasonable number of computable conjugates
of $\sigma_z^{(1)}$.

Even if the initial state is restricted to the deviation
$\sigma_z^{(1)}$, there are no known classical algorithms for
simulating an arbitrary sum of two bit hamiltonians as described
above.  Since many real-world situations involve highly mixed initial
states (e.g. most NMR experiments), such an algorithm is very
interesting. In fact there are ongoing experiments exploiting the fact
that many ${1\over r^3}$ interactions can be simulated in solid state
systems such as calcium-fluoride by modifying the dipolar interaction,
but are difficult to simulate using classical
computation~\cite{zhang:qc1998a}.
The observation that it it appears to be difficult to efficiently
implement such simulations in DQC1 makes this model useful and
the question of the relationship between the powers of DQC1 and DQCp or
classical computation non-trivial.

For any hamiltonian, one of its physical properties of interest is the
energy spectrum. This problem of determining the spectral density
function is not in the class of one-bit deterministic
functions. However, any decision problem based on the spectrum can be
thought of as a non-deterministic one-bit function. The
non-determinism is needed to permit the possibility that either
decision is acceptable. Without this possibility there can be
situations where we cannot gain arbitrarily high confidence in the
correctness of the decision.

In DQC1 it is possible to directly observe the spectrum with a
resolution inversely related to the effort used. Methods to accomplish
this type of observation are well known in the NMR literature, where
the natural hamiltonians are routinely modified using effective
hamiltonian techniques~\cite{waugh:qc1996a}. Decoupling methods used
to effectively remove interactions with some nuclei are an
application~\cite{waugh:qc1982a}. Here we give a simple method for
observing spectra from the point of view of quantum computation.

Let $U(t)=e^{-iHt}$ and assume that an efficient
quantum network for applying $U(t)$ with arbitrarily small error is
available. Note that the quantum network may increase in complexity if
less error is needed.  Given the quantum network for $U(t)$, we can
derive networks for applying $U(t)$ or $U^\dagger(t)$ to bits $2\ldots
n+1$, conditionally on the state of the first bit. Let $V(t)$ be the
unitary operator which maps $\ket{1}\ket{b}\rightarrow
\ket{1}U^\dagger(t/2)\ket{b}$ and $\ket{0}\ket{b}\rightarrow
\ket{0}U(t/2)\ket{b}$.  If we first apply a gate to transform the
input state $\sigma_z^{(1)}$ to $\sigma_x^{(1)}$, and apply
$V(t)$, then the deviation of the state becomes
\begin{eqnarray}
\rho_f &=& \sum_i (\cos(\lambda_i t)\sigma_x^{(1)}
                        + \sin(\lambda_i t)\sigma_y^{(1)})\ket{i}\bra{i}\\
  &=& {1\over 2^{n}}\sum_i (\cos(\lambda_i t)\sigma_x^{(1)}
                              + \sin(\lambda_i t)\sigma_y^{(1)}) + \mbox{rest}.
\nonumber
\end{eqnarray}
where the $\lambda_i$ and $\ket{i}$ are a complete set of eigenvalues
and corresponding eigenvectors (with repetition) of $H$. The second
identity gives the expansion of $\rho_f$ in terms of $\sigma_b$'s,
with terms not of interest suppressed.  The coefficients of $\sigma_x$
and $\sigma_y$ can be measured and the results combined into a single
complex number with value $f(t) = {1\over 2^{n+1}}\sum_i e^{-\lambda_i
t}$. This can be sampled at as many time points as desired and fourier
transformed to obtain a broadened energy spectrum. The same technique
can be used to measure the spectrum of a unitary operator
by restricting $t$ to integer multiples (corresponding to
powers of the operator). If the evolution
$V(t)$ is implemented as a power of $V(\Delta t)$, and the measurement
in DQC1 is non-destructive, then the spectrum can be observed directly
as is done in fourier transform NMR spectroscopy.


DQC1 is strictly less powerful then DQCp in the presence of oracles
(a.k.a.~black box).  Suppose that we are given a black box
implementing a unitary operator $U$ on our quantum system and we wish
to determine $\trace\bra{\bf 0}U^\dagger\sigma_z^{(1)}U\ket{\bf 0}$,
whose sign is the one bit answer computed by $U$ on input $\ket{\bf
0}$. One method for obtaining this answer using DQC1 involves
preparing a {\em pseudo-pure\/} state from the deviation
$\sigma^{(1)}_z$~\cite{cory:qc1997a,chuang:qc1997a}.  Such a state has
deviation proportional to that of a pure state. If the deviation is
proportional to that of $\ket{\bf 0}\bra{\bf 0}$, measurement of
$\sigma_z^{(1)}$ after applying $U$ is proportional to the desired
answer.

It turns out to be quite simple to prepare a kind of pseudo-pure state
from $\sigma^{(1)}_z$ using an ancilla bit, here labeled $0$.  Let
$T_n$ be the unitary operator mapping $\ket{b_0{\bf
0}}\rightarrow\ket{(b_0+1){\bf 0}}$ and for $b\not=0$,
$\ket{b_0b}\rightarrow\ket{b_0b}$ (addition of bits is modulo two). If
we first swap bits $0$ and $1$, then apply $T_n$, and finally flip bit
$0$, the deviation is given by $\sigma_z^{(0)}(2\ket{\bf 0}\bra{\bf
0}-I)$. If we apply $U$ to bits $1$ through $n$, then the coefficient
$\alpha$ of $\sigma_z^{(0)}\sigma_z^{(1)}$ in the state's deviation is
the answer we want. As discussed above, this coefficient can be
obtained using DQC1, with an intensity of $\alpha/2^{n+1}$. The
exponential loss of intensity implies that it is very difficult to
detect $\alpha$ above the noise.

Unfortunately, without the ability to analyze a specification of $U$
(for example, an implementation by a quantum network), we cannot do
much better, even if we know that the answer of $U$ is deterministic,
that is $\alpha \in \{-1,1\}$.  The most general form of a DQC1 algorithm
for determining the answer can be described by $k$ independent DQC1
computations consisting of quantum networks and calls to $U$ and an
inference function which attempts to determine $\alpha$ from the $k$
measurements.  Consider the first of these measurements. The
expectation of the result of the measurement can be written in the
form
\begin{equation}
v(U) = {1\over 2^m}\trace(V_r U \ldots U V_0\sigma_z^{(1)} V_0^\dagger U^\dagger \ldots U^\dagger
  V_r^\dagger\sigma^{(1)}_z),
\end{equation}
where $r$ is the number of invocations of $U$, the $V_i$ are the
quantum networks used in the computation using $m$ bits (of
which $m-n$ are ancillas).  Since $U$ is deterministic,
$U\ket{0}\ket{\bf 0} = \ket{b}\ket{\psi}$.  If $U$ is composed with
the operator $T$ which conditional on the state of bits $2$ to $n$
being $\ket{\psi}$ flips the first bit, we get an operator $U'$ which
is also deterministic but has the opposite answer.  We must be able to
distinguish between the two.  $T$ can be written in the form $T=I-2P$,
where $P$ is the pure state given by
${1\over 2}(I-\sigma_x^{1})\tensor\ket{\psi}\bra{\psi}$ (the $m-n$ ancillas
have been suppressed in this expression). For unitary operators $W_1$
and $W_2$ acting on $m\geq n$ bits, we have $|{1\over
2^m}\trace(W_1PW_2)|\leq {1\over 2^{n}}$.  By expanding $U'= U-2PU$ in
the expression for $v(U')$ we can write
\begin{eqnarray}
&v(U')& \\
  &=&{1\over 2^m}(\trace(V_r (-2P)U V_{r-1}U'\ldots) + \trace(V_r U V_{r-1} U'\ldots))\nonumber\\
      &=& {1\over 2^m}( \trace(V_r (-2P)U V_{r-1}U'\ldots) \nonumber\\&&+
                  \trace(V_r U V_{r-1} (-2P) U V_{r-2}U'\ldots) \nonumber\\
  &&+             \trace(V_r U V_{r-1} U V_{r-2} U'\ldots))\nonumber\\
      &=& {1\over 2^m}( \trace(V_r (-2P)U V_{r-1}U'\ldots) \nonumber\\
      && + 
                  \trace(V_r U V_{r-1} (-2P) U V_{r-2}U'\ldots) + \ldots + v(U)).\nonumber
\end{eqnarray}
Thus, $v(U') = {1\over 2^m}(a_1+\ldots+ a_{2r} + v(U))$, where
$|a_i|\leq {2^{m-n+1}}$.  It follows that $|v(U')-v(U)|\leq
{4r/2^n}$. It is clear that there is a sensitivity problem: To
confidently distinguish between $U'$ and $U$ requires exponentially
many experiments or invocations of the oracle.

The inequality derived in the previous paragraph suggests that it might
be possible to obtain the answer of $U$ with $O(2^n)$ invocations
of $U$ rather than the $O(2^{2n})$ experiments required to detect
a signal of strength ${1/2^n}$ in the presence of bounded noise.
We do not know whether this is possible.


We have introduced a new model of computation (DQC1) with power
between classical computation and deterministic quantum computation
with pure states (DQCp).  DQC1 is not as powerful as DQCp in the
presence of unitary oracles but can solve problems in physics
simulation for which there are no known efficient classical
algorithms. Since DQC1 requires only one quantum bit of information,
the possibility that it adds power to classical computation is
surprising. If this is the case, the usual reasons given for why
quantum computation appears to be so powerful may have to be revised.
On the other hand, a proof that DQC1 can be efficiently simulated by
classical computation would be extremely interesting, as this could
lead to practical algorithms for simulations of many experimentally
interesting physical situations.

\bigskip
{\bf Acknowledgments.} The reader may be interested to learn that the
authors of this letter vehemently disagree on the likelihood that DQC1
is strictly more powerful than classical computation. We benefited
greatly from discussions with David Cory and Tim Havel, who introduced
us to the exciting world of solid state NMR. Michael Nielsen's
encouragement was quintessential in pushing us to write up our ideas.
We thank the National Security Agency for financial support.


\end{document}